\documentclass[RNAAS]{aastex63}
\usepackage{amsmath}

\newcommand{\mlt}{$\alpha_{\text{MLT}}$}
\usepackage{CJK}
\begin{document}

\title{Revised Best Estimates for the Age and Mass of the Methuselah Star HD 140283 using MESA and Interferometry and Implications for 1D Convection }

\correspondingauthor{Meridith Joyce}
\email{Meridith.Joyce@anu.edu.au}
\author[0000-0001-5752-3847]{Jianling Tang\begin{CJK*}{UTF8}{gbsn} (汤健羚)\end{CJK*}}

\affiliation{Research School of Astronomy $\&$ Astrophysics, Australian National University, Canberra, ACT 2611, Australia}

\author[0000-0002-8717-127X]{Meridith Joyce}
\affiliation{Research School of Astronomy $\&$ Astrophysics, Australian National University, Canberra, ACT 2611, Australia}
\affiliation{ARC Centre of Excellence for All Sky Astrophysics in 3 Dimensions (ASTRO 3D)}


\begin{abstract}

In light of recently revised observational measurements of the radius and spectroscopic parameters of the extremely old and metal-poor Gaia benchmark star HD 140283---also known as the Methuselah star due to prior suggestions that its age is in tension with the age of the Universe---we present new, best estimates for the star's mass and age from stellar modeling. These are derived using 1D stellar evolutionary tracks computed with MESA and the most up-to-date measurements from CHARA interferometry. Excluding modeling variance from the uncertainties, we report a mass of $0.809 \pm 0.001 M_{\odot}$ and an age of $12.01 \pm 0.05$ Gyr ($1 \sigma$). When dominant sources of modeling uncertainty are taken into account, we report $0.81 \pm 0.05 M_{\odot}$ and $12 \pm 0.5$ Gyr, respectively. These results are consistent with recent literature, and the best-fitting age is not in conflict with the currently accepted age of the universe ($13.5$ Gyr; \citealt{PlanckCollab}).
\end{abstract}

\section{Introduction}
Due to its old age, close proximity, and very low metallicity ([Fe/H] = $ -2.29 ^{\pm 0.10}_{\pm 0.04}$ dex), 
the Methuselah star's importance as a benchmark has been well established over the past two decades \citep{VandenBerg2000,VandenBerg2002,2013AAS...22144308B,VandenBerg2014,VandenBerg2016,Creevey2015,Joyce2018a,Jimenez2019}.  
With the newest radial measurements obtained from the interferometric instrument PAVO at the CHARA array \citep{Karovicova2020}, it is prudent to revise our predictions for HD 140283's other fundamental stellar parameters accordingly.

\section{METHODS}
We use the Modules for Experiments in Stellar Astrophysics (MESA v11701; \citealt{MESAI, MESAII, MESAIII, MESAIV, MESAV}) software and an adaptive grid searching method \citep{Joyce2018a, Joyce2018b, Murphy2021} to construct a set of metal-poor stellar evolutionary tracks for HD 140283. 
We vary as inputs the
mass, initial composition, and convective mixing length, \mlt, which parameterizes the efficiency of energy transport by convection in the low-mass star's outer envelope according to the
Mixing Length Theory (MLT) formalism first established by \citet{BohmVitense1960}. 

Our search covers masses from 0.6 to 2.0 $M_{\odot}$, 
$\alpha_{\text{MLT}}$ values $1.5-2.0\times$ the pressure scale height, $H_p =d\ln P/d\ln T$, and
input metallicities, 
($Z_{\text{in}}$), between 0.00006 and 0.002 ([Fe/H] $= -2.46$ to $-0.94$).

%
Since previously reported ages for HD 140283 approach the age of the universe (e.g., \citealt{VandenBerg_2000}), we uniformly adopt the helium abundance of the early universe, $Y_\text{in} = Y_{\text{primordial}} = 0.245$.

\begin{figure}[htp] 
    \hskip-35pt
    \includegraphics[scale=0.72]{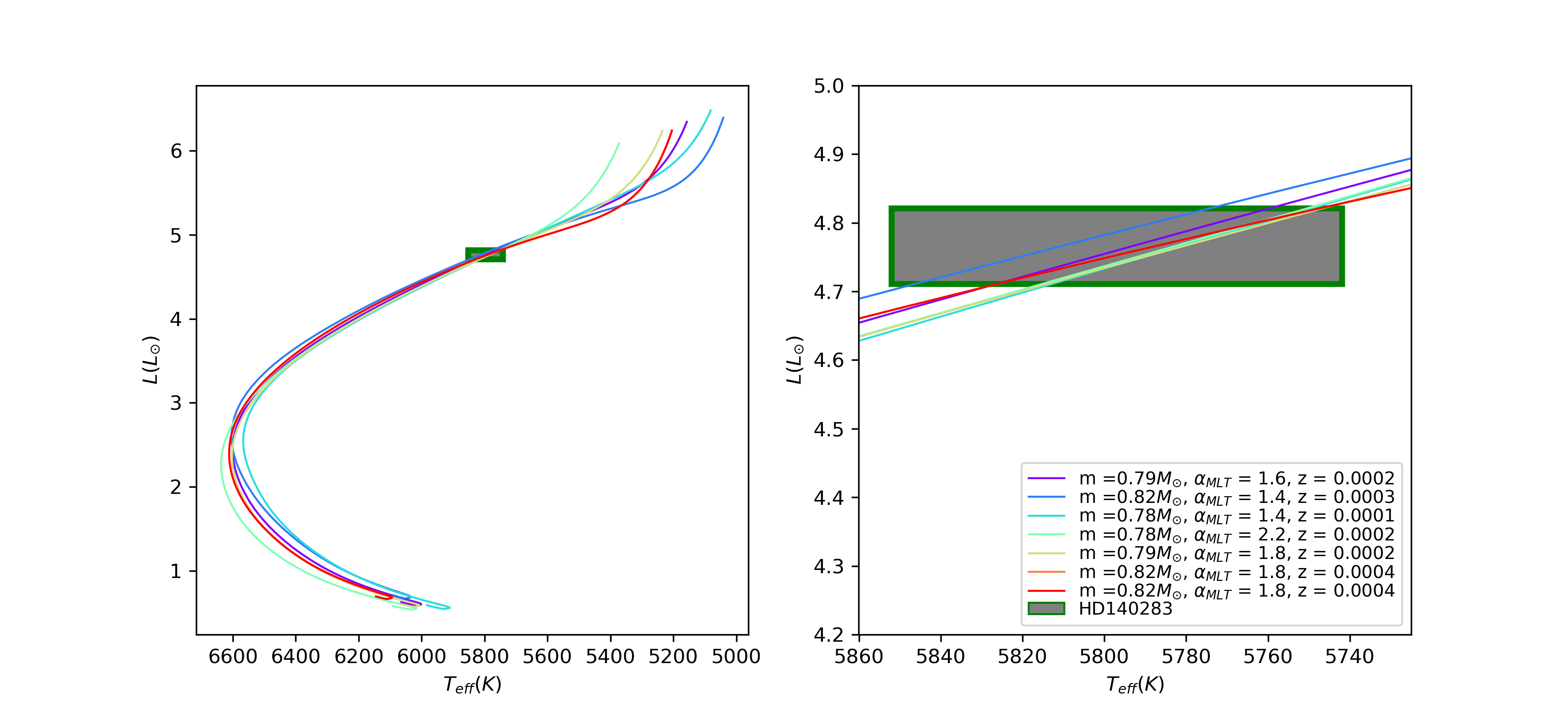}
    \caption{Tracks with varying masses, mixing lengths,
    and metallicities on the HR diagram. Tracks intersecting \citet{Karovicova2020}'s constraints are considered valid. Selected tracks are chosen according to their minimization of equation \ref{eqn:cost}.
    }
    \label{fig:tracks}
\end{figure}

Models are considered ``valid'' according to agreement with the effective temperature and luminosity constraints quoted in \citet{Karovicova2020} and shown in Figure \ref{fig:tracks}. A pseudo-$\chi^2$ ranking scheme assesses relative goodness-of-fit among all classically valid tracks. The pseudo-$\chi^2$ cost function, equation~\eqref{eqn:cost}, assigns equal statistical weighting to the modeled radius ($R$), luminosity ($L$), and surface metal-to-hydrogen mass fraction
$Z/X_{\text{surf}}$---a metric of the form introduced in \citep{Joyce2018b}. To wit:%
\begin{equation}
\catcode`&=9
    \begin{aligned} 3s^{2}=&\left[\frac{R{_\mathrm{obs}}-R_{ \mathrm{mod}}}{\sigma_{R, \text { obs }}}\right]^{2} &+\left[\frac{L{_\mathrm{obs}}-L_{ \mathrm{mod}}}{\sigma_{L, \text { obs }}}\right]^{2} &+\left[\frac{Z/X_{\mathrm{obs}}-{Z/X_\mathrm{mod}}}{\sigma_{Z/X,{\mathrm{obs}}}}\right]^{2}\end{aligned},
    \label{eqn:cost}
\end{equation}
where $R_{\text {obs}}$ is the interferometric radius of HD 140283 with uncertainty $\sigma_{R,\text{obs}}$, $R_{\mathrm{mod}}$ is the modeled radius, and similarly for $L$ and $Z/X$.

\section{RESULTS}
The model that minimizes equation \eqref{eqn:cost} has a mass of 0.79 $M_{\odot}$, $Z_i$ of 0.0002, and \mlt$=1.6$ $\mathrm{H_p}$. 
However, as many models yield $\chi^2 \le 1.0$, the optimal parameters are more accurately determined by measuring the frequencies of values among all acceptable models.
Because the mass and age estimates do not follow a normal distribution, the confidence interval cannot be calculated directly from the standard deviations. For large data sets, the sampling distribution of the studentized mean is approximately normal \citep{cook_weisberg_1975}. Consequently, $\frac{\bar{x}-\mu}{s / \sqrt{n}} \approx \mathrm{N}(0,1)$ is used to construct the studentized error, representing a confidence interval
of 95\%.

We apply frequency statistics to parameters from all points of observational intersection: 198 tracks of 1659 total intersect at one or more timesteps (${\sim} 12$\%).
This yields a $1 \sigma$ mass and age for HD 140283 of $0.809 \pm 0.001 M_{\odot}$ and $12.01 \pm 0.05$ Gyr, respectively. 

As this analysis uses invariant physical assumptions,
the uncertainties do not include contributions from modeling systematics. To provide more realistic estimates of global uncertainty, 
we replicate the numerical experiment using two different prescriptions for atmospheric surface boundary conditions: 
\begin{enumerate}
\item  pre-computed photosphere tables based on \citet{1999ApJ...512..377H,1999ApJ...525..871H}'s PHOENIX models and \citet{2003IAUS..210...45K}); and
\item Eddington $t-\tau$ integration \citep{1930MNRAS..90..279E};
\end{enumerate}
\noindent likewise, two convective overshoot prescriptions: MESA's default  \verb|`step'| method and the \verb|`exponential'| overshooting scheme described by \citet{2000A&A...360..952H}.
These physics, along with heavy element diffusion, were identified as large sources of instrumental variance in the Dartmouth Stellar Evolution Program (DSEP) \citep{Joyce2018a}, MESA (Joyce et al., 2021, in prep), and in other stellar evolution codes (\citealt{Lebreton2014, Tanner2014}).  
Presently, we relegate diffusion considerations to future work but strongly emphasize their importance. Results from tests across physical prescriptions revise our mass and age for HD 140283 to $0.81 \pm 0.05 M_{\odot}$ and $12 \pm 0.5$ Gyr, respectively.

\section{DISCUSSION}
Recently, 
\citet{Creevey2015} reported a best fitting mass and age of  $0.780 \pm  0.010 \ M_{\odot}$ and $13.7 \pm 0.7 \text { Gyr }\left(A_{V}=0.0 \ \mathrm{mag}\right)$. 
%
%
\citet{Joyce2018a} found a best-fitting mass range of 0.74--0.79 $M_{\odot}$ 
and a mass-dependent age range of 12.5 to 14.9 Gyr, with higher masses corresponding to lower ages. \cite{Karovicova2020} reported a best fitting mass of $0.77 \pm 0.03 M_{\odot}$.
Our median mass and age are slightly higher and lower, respectively, than the results cited above but remain firmly consistent with recent literature and not in conflict with the universe's age.

Importantly, our models show sensitivity to the mixing length parameter, preferring values between $1.6$--$1.8 H_p$. By comparing the models' preferred $\alpha_{\text{MLT}}$ against solar mixing length calibrations calculated under associated physical conditions, we find that HD 140283 requires a convective mixing length $10$--$20\%$ below the solar value. This result contributes to an increasing body of literature showing significantly sub-solar  $\alpha_{\text{MLT}}$ values are necessary to reproduce both the observed properties of metal-poor stars in general (e.g., \citealt{GuentherDemarque2000, Tayar2017, Viani2018}) and of HD 140283 specifically (\citealt{Creevey2015, Joyce2018a}). 
%


\bibliography{RNAAS.bib}{}
\bibliographystyle{aasjournal}

\end{document}